\documentclass[useAMS,usenatbib,a4paper]{mn2e}
\usepackage{graphicx}
\usepackage{amsmath}
\usepackage{txfonts}
\usepackage{mathrsfs}
\usepackage{color}
\usepackage{ulem}   

\usepackage{hyperref}
\hypersetup{hypertex=true,
	colorlinks=true,
	linkcolor=blue,
	anchorcolor=blue,
	citecolor=blue}

\newcommand{\be}{\begin{equation}}
	\newcommand{\ee}{\end{equation}}
\newcommand{\bea}{\begin{eqnarray}}
	\newcommand{\eea}{\end{eqnarray}}

\title[Cosmic Distance Duality Relation]{Test of the cosmic distance duality 
	relation for arbitrary spatial curvature}
\author[Qin, Melia \& Zhang]{Jin Qin$^{1}$\thanks{E-mail: 201821160009@mail.bnu.edu.cn},
	{Fulvio Melia$^2$}\thanks{E-mail: fmelia@email.arizona.edu. John Woodruff Simpson Fellow}
	and {Tong-Jie Zhang$^1$\thanks{E-mail: tjzhang@bnu.edu.cn}} \\
	$^1$Department of Astronomy, Beijing Normal University, Beijing 100875, China\\
	$^2$Department of Physics, The Applied Math Program, and Department of 
	Astronomy,
	The University of Arizona, AZ 85721, USA
}

\begin{document}
	
	\date{October 6, 2020}
	
	\pagerange{\pageref{firstpage}--\pageref{lastpage}} \pubyear{2020}
	\maketitle
	
	\label{firstpage}
	
	\begin{abstract}
		The cosmic distance duality relation (CDDR), $ \eta(z)=(1+z)^2d_A(z)/d_L(z) = 1$, 
		is one of the most fundamental and crucial formulae in cosmology. This relation 
		couples the luminosity and angular diameter distances, two of the most often
		used measures of structure in the Universe. We here propose a new 
			model-independent method to test this relation, using strong gravitational lensing 
			(SGL) and the high-redshift quasar Hubble diagram reconstructed with a 
			B{\'e}zier parametric fit. We carry out this test without pre-assuming a zero spatial curvature, adopting instead the value $\Omega_K=0.001\pm 0.002$ optimized by 
			{\it Planck} in order to improve the reliability of our result. We parametrize 
		the CDDR using $\eta(z)=1+\eta_0 z$, $1+\eta_1 z+\eta_2 z^2$ and $1+\eta_3 z/(1+z)$, 
		and consider both the SIS and non-SIS lens models for the strong lensing. 
		Our best fit results are: $\eta_0=-0.021^{+0.068}_{-0.048}$, 
			$\eta_1=-0.404^{+0.123}_{-0.090}$, $\eta_2=0.106^{+0.028}_{-0.034}$, and
			$\eta_3=-0.507^{+0.193}_{-0.133}$ for the SIS model, and 
			$\eta_0=-0.109^{+0.044}_{-0.031}$ for the non-SIS model. The measured 
			$\eta(z)$, based on the {\it Planck} parameter $\Omega_K$, is essentially
                        consistent with the value ($=1$) expected 
			if the CDDR were fully respected. For the sake of comparison,
			we also carry out the test for other values of $\Omega_K$, but find that
			deviations of spatial flatness beyond the {\it Planck} optimization are in
			even greater tension with the CDDR. Future measurements of SGL may improve 
			the statistics and alter this result but, as of now, we conclude that the
			CDDR favours a flat Universe.
	\end{abstract}
	
	\section{Introduction}
	In cosmology, the luminosity distance, $D_L$, and angular diameter distance, $D_A$, are widely 
	used measures for many astronomical observations. Etherington \citep{Etherington_1993} first
	argued for a simple, yet profound, relationship between them,
	\be
	\eta(z)\equiv \frac{D_A}{D_L}(1+z)^2=1\;,
	\ee
	often called the Etherington distance duality relation (DDR) 
	\citep{Etherington_2007}, or the cosmic distance duality relation (CDDR). There are only 
	three basic conditions required to make this relationship work:
	
	\begin{itemize}
		\item[1.] The spacetime is described by a metric theory of gravity,
		\item[2.] The photons travel along null geodesics,
		\item[3.] The number of photons is conserved.
	\end{itemize}
	
	One may test the validity of the CDDR for any given cosmological model. An observed
	deviation from the CDDR may imply some dramatic new physics. One should keep in mind,
	however, that unrecognized systematic uncertainties in the observations may also lead 
	to a breakdown of the CDDR. It is therefore essential to select a means of testing 
	the Etherington relation that is as accurate and reliable as possible.
	
	An early attempt to validate the CDDR was presented by \cite{Bassett_2004_cddbeigins},
	who combined Type Ia SN data (SNe Ia) to measure $D_L$, and radio galaxies, compact 
	radio sources and X-ray clusters to measure $D_A$. Several other early studies used a 
	similar strategy to validate the CDDR based on the same types of data 
	\citep{Uzan_2004_prd_cdd,Bernardis_2006,Khedekar_2011_PRL_cdd21cm,Holanda_2010_APJL,
		Li_2011_APJL_cdd,Nair_2011_cdd,Meng_2012_cdd,Ellis_2013}.
	To obtain $D_L$ and $D_A$, however, these investigations had to adopt a specific 
	cosmological model, usually flat $\Lambda$CDM. But if the background model is incorrect,
	or imperfect, the result is less credible. More recent work has focused on testing 
	the CDDR in a model-independent way \citep{Liao_2016ApJ_cdd, Lv_2016PDU_cdd, 
		Li_2018_MNRAS_cdd, Lin_2018_MNRAS_cdd, Ruan_2018_APJ_cdd, Lyu_2020_apj_cdd}. Interestingly, 
	none of these have indicated any significant violation of the CDDR within the 
	margin of testing uncertainties. Nevertheless, all these works assumed a spatially flat 
	cosmic background, so the conclusion inferred thus far may be biased.
	
	In this paper, we propose a new model-independent method to test the CDDR without 
	assuming a spatially flat Universe. We use the recently released 161 galaxy-scale 
	strong gravitational lensing (SGL) systems \citep{Chen_2019MNRAS_NewLensData} to measure 
	$D_A$. These are then combined with observations of high redshift quasars, 
		which we use to infer $D_L$ via a B$\mathrm{\acute{e}}$zier parametric fit, to test 
		the CDDR at relatively high precision. Below, we shall describe both SIS and non-SIS 
	lens models used for the SGL, along with an evaluation of their influence on various 
	forms of parameterization for $\eta(z)$. To gauge the impact of spatial
		curvature on the CDDR, we test not only the {\it Planck} optimization of $\Omega_K$
		\citep{Planck_2018}, but also 20 other assumed values.
	
	In \S~2, we introduce the methodology we shall follow to test the CDDR 
		for a Universe with arbitrary curvature. Then we describe the observations of SGL's 
		and the reconstruction of the luminosity distance using high-redshift quasars in 
		\S~3. We present our results in \S~4, and end with our conclusions in \S~5.
	
	\section{Methodology}
	A homogeneous, isotropic expanding Universe can be characterized by the 
	Friedmann-Lema\^itre-Robertson-Walker (FLRW) metric,
	\be
	ds^2=-c^2 dt^2 +a^2(t)\left(\frac{dr^2}{1-Kr^2}+r^2d\Omega^2\right)\;,
	\label{FLRW}
	\ee
	where $a(t)$ is the scale factor, $c$ is the speed of light, and the constant 
	$K$ represents the spatial curvature. Using Equation~\eqref{FLRW}, we can define 
	a dimensionless distance between the source at redshift $z_s$ and the lens at $z_l$
	\citep{Rasanen_2015PRL}:
	
	\be
	d(z_l,z_s)\equiv {(1+z_s)\over D_H}D_A(z_l,z_s)\;,
	\ee
	where $D_H\equiv c/H_0$. Thus, 
	\be
	d(z_l,z_s)=\frac{1}{\sqrt{\Omega_K}}\textup{sinh}\left(\sqrt{\Omega_K}\int^{z_s}_{z_l}
	\frac{H_0}{H(z)}dz\right)\;,
	\label{dls}
	\ee
	where $D_A(z_l,z_s)$ is the angular diameter distance between $z_s$ and $z_l$,
	and $\Omega_K\equiv -K/H_0^2$ which, according to {\it Planck} \citep{Planck_2018}, 
	is a small positive number. For convenience, we shall use the following notation: 
	$d_{ls}\equiv d(z_l,z_s)$, $d_{l}\equiv d(0,z_l)$ and $d_{s}\equiv d(0,z_s)$. 
	
	Using Equation~\eqref{dls}, the relationship between $d_{ls}$, $d_l$ and $d_s$ may
	be written 
	\be
	d_{ls}=d_s\sqrt{1+\Omega_K d_l^2}-d_l\sqrt{1+\Omega_Kd_s^2}\;,
	\label{sumrule}
	\ee
	which is usually referred to as the distance sum rule 
	\citep{peebles1993principles,Bernstein_2006,Rasanen_2015PRL}. If the Universe has 
	positive spatial curvature, as suggested by {\it Planck}, then the distance 
	$d_{ls}$ will be smaller than the difference between $d_s$ and $d_l$. 
	
	To facilitate the use of strong lensing systems in order to find the angular diameter
	distance via Equation~\eqref{sumrule}, we rewrite this equation as follows: 
	\be
	\frac{d_{ls}}{d_{s}}=\sqrt{1+\Omega_K 
		d_l^2}-\frac{d_l}{d_s}\sqrt{1+\Omega_Kd_s^2}\;,
	\label{sumrule1}
	\ee
	where the ratio $d_{ls}/d_s$ is extracted directly from the SGL's. Using the 
	relations $d_l=(1+z_l) D_A(0,z_l)/D_H$ and $d_s=(1+z_s)D_A(0,z_s)/D_H$, and 
	expressing $D_A(z_l,z_s)$, $D_A(0,z_l)$ and $D_A(0,z_s)$ as $d_A^{\,ls}D_H$, 
	$d_A^{\,l}D_H$, and $d_A^sD_H$, respectively, we then find from Equation~\eqref{sumrule1} that
	\be
	\begin{split}
		\frac{d_A^{\,ls}}{d_A^{\,s}}&=\sqrt{1+\Omega_K(1+z_l)^2 {d_A^{\,l}}^2}\\
		&\qquad -\sqrt{1+\Omega_K(1+z_s)^2 
			{d_A^{\,s}}^2}\frac{(1+z_l)\,d_A^{\,l}}{(1+z_s)\,d_A^{\,s}}\;.\qquad
	\end{split}
	\ee
	Note that $d_A^{\,ls}$, $d_A^{\,l}$ and $d_A^s$ are dimensionless. Then, writing the CDDR as
	\be
	\eta(z)=\frac{d_A}{d_L}(1+z)^2\;,
	\ee
	we find that 
	\be
	\frac{d_A^{\,ls}}{d_A^{\,s}}\equiv{\mathscr{R}}\left(d_L^{\,l},d_L^s,z_l,z_s\right)=K_l-\Phi K_s\;,
	\label{dadl}
	\ee
	where 
	\begin{subequations}
			\begin{align}
				&\Phi\equiv\frac{d_L^{\,l}\,\eta(z_l)(1+z_s)}{d_L^{\,s}\,\eta(z_s)(1+z_l)}\label{a}\;,\\
				&K_l\equiv\sqrt{1+\left(\frac{d_L^{\,l}\,\eta(z_l)}{1+z_l}\right)^2 \Omega_K}\label{b}\;,\\
				&K_s\equiv\sqrt{1+\left(\frac{d_L^{\,s},\eta(z_s)}{1+z_s}\right)^2\Omega_K }\label{c}\;.
			\end{align}
	\end{subequations}
	The quantities $d_L^{\,l}$ and $d_L^{\,s}$ are the dimensionless luminosity distance at 
	$z_l$ and $z_s$, respectively. In this paper, the ratio $d_A^{\,ls}/{d_A^s}$ is 
		obtained from the SGL's, while $d_L^{\,l}$ and  $d_L^{\,s}$ are inferred from the 
		quasar data.
	
	We shall use three types of parameterization for $\eta(z)$:
	\begin{subequations}
		\begin{align}
			&\eta(z)=1+\eta_0 z\;,\label{a1}\\
			&\eta(z)=1+\eta_1 z+\eta_2 z^2\;, \label{b1}\\
			&\eta(z)=1+\eta_3\frac{z}{1+z}\;.\label{c1}
		\end{align}
	\end{subequations}
	Combining Equations~\eqref{dadl} and \eqref{a1}--\eqref{c1}, we 
	may then optimize the CDDR by maximizing the likelihood function,
	\be
	{\mathscr{L}}=\prod_{i=1}^{n}\frac{1}{\sqrt{2\pi}\ 
		\sigma_i}\exp\left(-\frac{\left( 
		{d_A^{\,ls}}_i/{d_A^{\,s}}_i-\mathscr{R}({d_L^{\,l}}_i,
		{d_L^{\,s}}_i,{z_l}_i,{z_s}_i)\right)^2}{2\sigma_i^2}\right)\;,
	\label{likehood}
	\ee
	where $n$ is the number of lenses, ${\mathscr{R}}({d_L^{\,l}}_i, {d_L^{\,s}}_i, 
	{z_l}_i, {z_s}_i) $ is given by Equation~\eqref{dadl}, and the variance can 
	be written 
	\be
	\sigma_i^2=\sigma_{int}^2+\sigma_{A,i}^2+\sigma_{{\mathscr{R}},i}^2\;.
	\ee
	In this expression, $\sigma_{int}$ is the global intrinsic dispersion, 
	$\sigma_{A,i}$ is the measurement uncertainty in ${d_A^{\,ls}}_i/{d_A^{\,s}}_i$,
	and $\sigma_{{\mathscr{R}},i}$ is the uncertainty in 
	${\mathscr{R}}({d_L^{\,l}}_i,{d_L^{\,s}}_i,z_l,z_s)$ ,which can be derived using
	an error propagation formula.
	
	\section{Data}
	\subsection{Angular Diameter Distance Obtained From SGL's}
	SGL's are widely used in constraining cosmic parameters, as well as testing the geometry 
	of the universe. In general, there is a relationship between $d_A^{\,ls}/d_A^{\,s}$ and 
	parameters that characterize the SGL: 
	\be
	\frac{d_A^{\,ls}}{d_A^{\,s}}=\frac{c^2\theta_{ E}\,{\mathscr{F}}(\gamma,\delta,\beta)}{2\sqrt{\pi}\,
		\sigma_0^2}\left(\frac{\theta_{\rm eff}}{2\theta_E}\right)^{2-\gamma}\;,
	\label{D_Als}
	\ee 
	where $\theta_{E}$ is the Einstein radius and
	\be
	\sigma_0=\sigma_{ap}\left(\frac{\theta_{\rm eff}}{2\theta_{ap}} \right)^\nu 
	\ee
	is the velocity dispersion. This latter equation is known as the aperture
	correction formula, written in terms of the ``aperture'' velocity dispersion,
	$\sigma_{ap}$. Its value, and that of $\theta_{\rm eff}$, may be found in
	\cite{Chen_2019MNRAS_NewLensData}. As we shall see in Table~\ref{table2} below, the
	``correction'' index $\nu$ is much smaller than $1$. The function
	${\mathscr{F}}(\gamma,\delta,\beta)$ is defined as follows:
	\be
	\begin{split}
		\mathscr{F}(\gamma,\delta,\beta)&=\frac{\Gamma \left(\frac{\gamma }{2}\right) 
			\Gamma \left(\frac{\delta }{2}\right) \Gamma \left(\frac{1}{2} (\gamma +\delta 
			-5)\right)}{2 \Gamma \left(\frac{\gamma -1}{2}\right) \Gamma \left(\frac{\delta 
				-3}{2}\right) \Gamma \left(\frac{\gamma +\delta }{2}\right)}\\
		&\qquad\times\frac{-\beta(\gamma +\delta -3)+\gamma +\delta -2}{-2 \beta 
			+\gamma +\delta -2}\;,\qquad
	\end{split}
	\ee
	where $\gamma$ is related to the total mass density, $\delta$ is related to the
	luminosity density and $\beta$ is the stellar orbital anisotropy 
	\citep{Chen_2019MNRAS_NewLensData,Lyu_2020_apj_cdd}. 
	
	\begin{figure}
		\centering
		\includegraphics[width=0.45\textwidth]{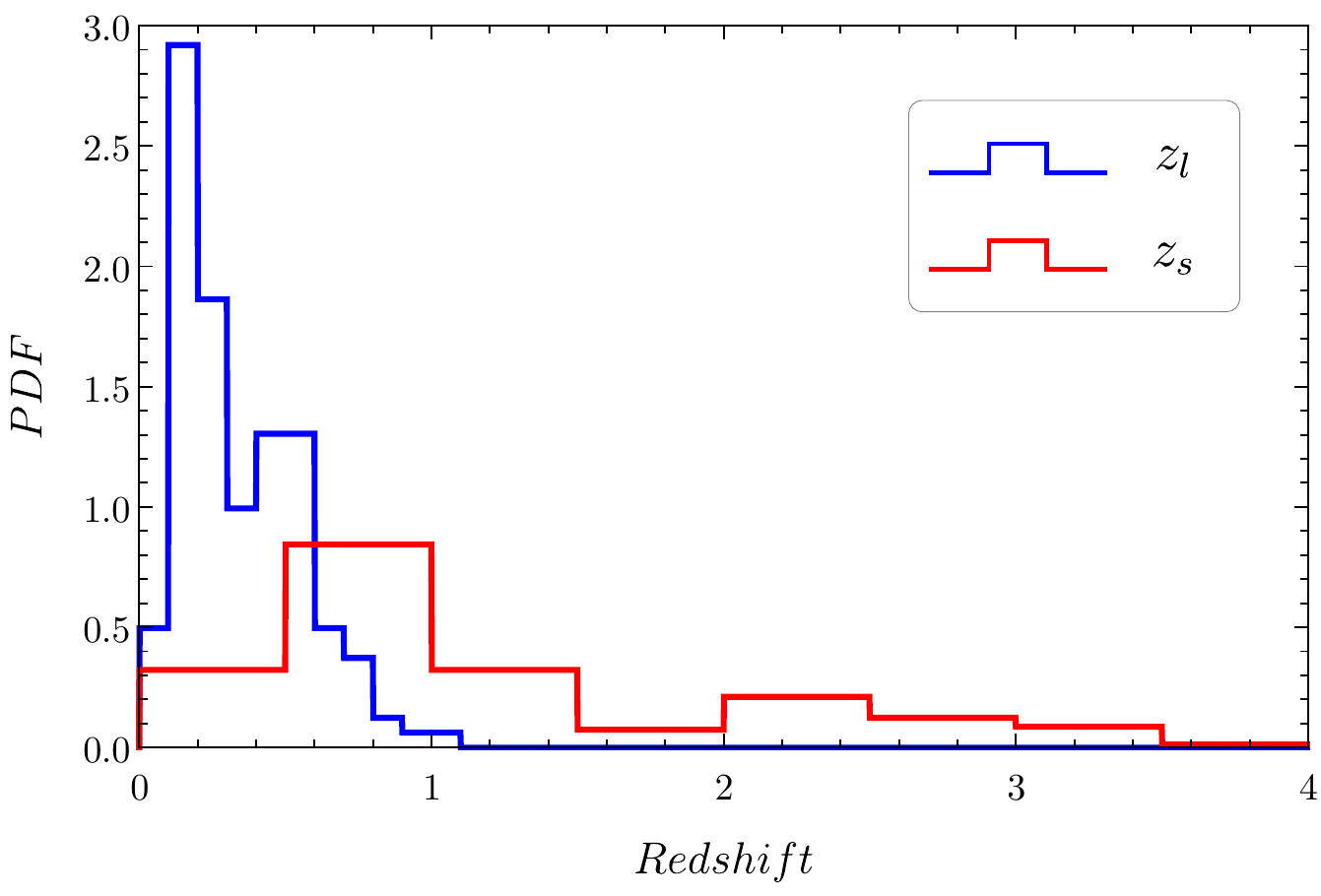}
		\caption{The redshift probability density histogram of lenses and sources 
			for the 161 recently reported SGL's \citep{Chen_2019MNRAS_NewLensData}.}
		\label{pdf}
	\end{figure}
	
	Combing this relation and the error propagation formula, we also obtain the uncertainty 
	in the SGL measurements for the non-SIS model:
	\be
	\sigma_A=\frac{d^{\,ls}_A}{d^{\,s}_A}\sqrt{(1-\gamma )^2
		\left( \frac{ \sigma _{\theta_{\rm E}}}{\theta_{\rm E}}\right)^2 +
		4\left( \frac{\sigma_{\sigma _0}}{\sigma_0}\right)^2}\;, 
	\ee
	where $\sigma_{\sigma_0}$ is the uncertainty in the velocity dispersion $\sigma_0$. 
	In this paper we assume that $\theta_{\rm E}$ has a flat uncertainty of five percent, 
	i.e., $\sigma_{\theta_{\rm E}}=0.05\,\theta_{\rm E}$. 
	
	When using the singular isothermal sphere (SIS) model, i.e. $\gamma=2$, $\delta=2$, 
	$\beta=0$, $\mathscr{F}$ reduces to $1/(2\sqrt{\pi}) $ and Equation~\eqref{D_Als} 
	simplifies to 
	\be
	\frac{d_A^{\,ls}}{d_A^{\,s}}=\frac{c^2\theta_{\rm E}}{4\pi\,\sigma_{SIS}^2}\;,
	\ee
	and the uncertainty is 
	\be
	\mathrm{
		\sigma_{A-SIS}=\frac{c^2\theta_{\rm E}}{4\pi\,\sigma_{SIS}^2}
		\sqrt{\left(\frac{\sigma_{\theta_{\rm E}}}{\theta_{\rm E}}\right)^2+ 
			4\left( \frac{\sigma_{\sigma_{SIS}}}{\sigma_{SIS}}\right)^2 }\;,
	}
	\ee
	where $\sigma_{SIS}\equiv f_e \sigma_0$, and $f_e$ corrects the deviation from $\sigma_0$.
	For the SIS model, intermediate-mass elliptical galaxies ($\mathrm{200\ km\ s^{-1} < \sigma_{\it ap} \le 300\ km\ s^{-1}}$) are the most reliable for our test \citep{cao_2016_mnras}. 
		For the non-SIS model, we use three subsamples (SLACS, S4TM, SL2S, {\color{blue}totaling} 121 quasars) to reduce the
		systematic errors, {\color{blue}taken from 161 SGL systems} (see fig.~\ref{pdf}). The subsamples bias the 
		SIS model slightly so, among the 161 SGL's, we keep only those with the limit of 
		$\sigma_{ap}$ mentioned above, leaving 109 lenses for the SIS model.
	
	\subsection{Luminosity Distance Obtained From High Redshift Quasars}
	Although Type Ia SNe are widely used standard candles to determine the luminosity 
		distance, the limitations of these sources are now quite apparent. Their first disadvantage is 
		observable events are restricted to redshifts $\lesssim 2$ \citep{Jones_2013}, so one cannot
		use them to measure $D_L$ at high $z$. In addition, one must optimize several so-called 
		``nuisance" parameters for their lightcurve simultaneously with other parameters in the
		specified cosmology. These ``nuisance" parameters have very different values for different 
		models, so the luminosity distance obtained from SNe is very model-dependent. A recently
		compiled high-quality catalog of high-$z$ quasars largely circumvents such problems
                \citep{Risaliti_2019_Quasars}.
		
		Quasars are highly luminous objects visible at redshifts sometimes exceeding $\sim 7$. 
		It has been known for several decades (see also \citealt{Fulvio_2019_quasar}) that their 
		UV and X-ray emissions are correlated with a simple log-linear relation
		\begin{equation}
			\log \left(L_{\mathrm{X}}\right)=\varepsilon \log \left(L_{\mathrm{UV}}\right)+\alpha\;,
			\label{quasar1}
		\end{equation}
		where $L_\mathrm{X}$ and $L_{\mathrm{UV}}$ are the rest-frame monochromatic luminosities at 2 
		keV and 2,500 $\AA$, respectively, and $\varepsilon$ and $\alpha$ are two constants. In order 
		to obtain the luminosity distance, we rewrite equation \eqref{quasar1} as 
		\begin{equation}
			\log _{10} D_L=\frac{1}{2(\varepsilon-1)}\left( \mathrm{log} _{10} 
			F_{\mathrm{X}} -\varepsilon \log _{10} F_{\mathrm{UV}}-\tilde{\alpha}\right)\;,
			\label{quasarlumi}
		\end{equation}
		where $\tilde{\alpha} \equiv \alpha+(\varepsilon-1)\mathrm{log}_{10}4\pi$. To fit the data 
		more conveniently, we write the distance modulus as 
		\begin{equation}
			\begin{split}
				\mu(z)&=\frac{5}{2(\varepsilon-1)}\left( \mathrm{log} _{10} F_{\mathrm{X}} -\varepsilon \log _{10} F_{\mathrm{UV}}-\tilde{\alpha}\right)+\\
				&\qquad 5\mathrm{log}_{10}\left(\frac{\mathrm{cm}}{\mathrm{Mpc}}\right) +25\;.
				\label{distance modulus}
			\end{split}
		\end{equation}
		
		\begin{figure}
			\centering
			\includegraphics[width=0.45\textwidth]{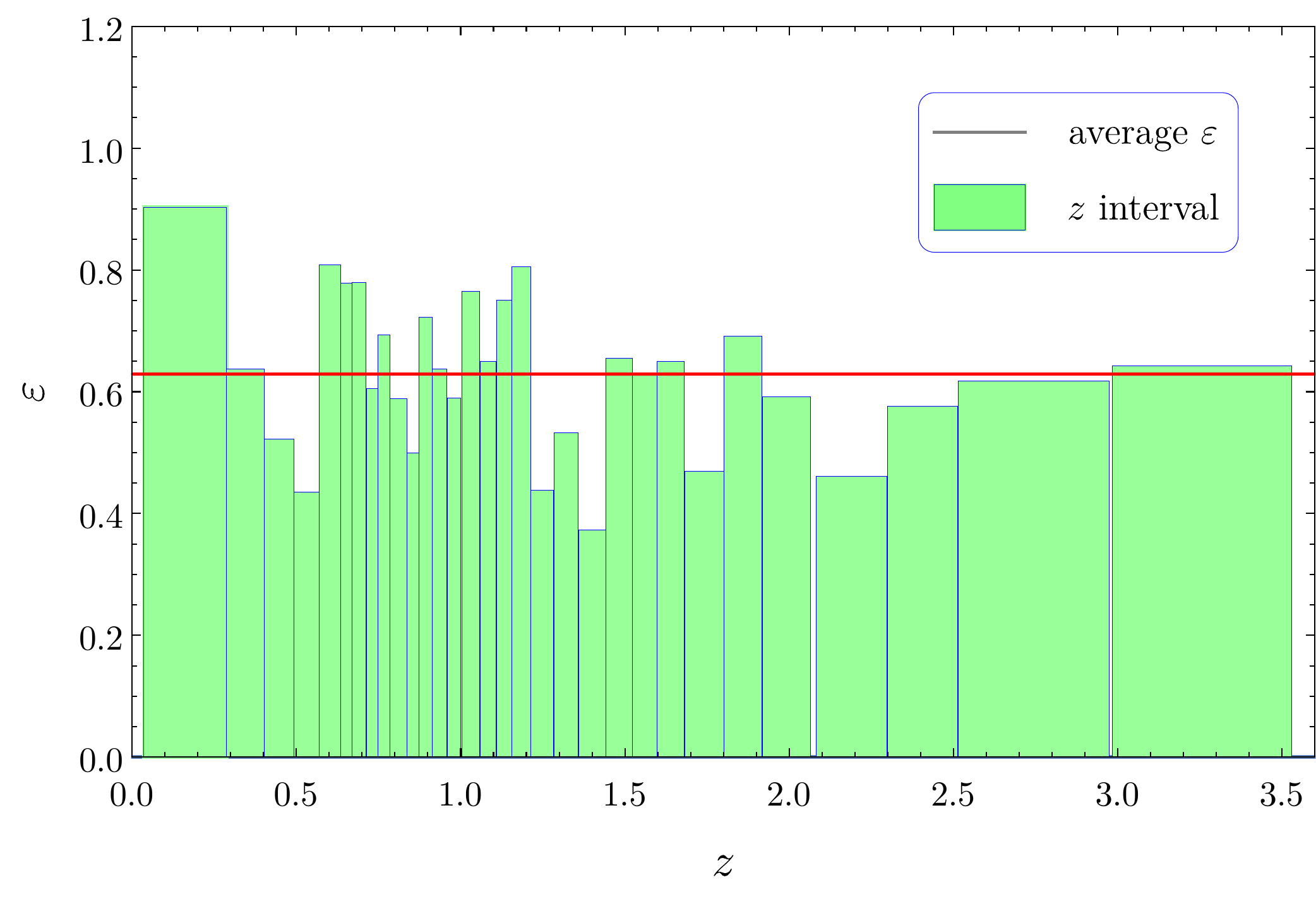}
			\caption{Determination of the slope $\varepsilon$. The width of each green rectangle 
				is chosen to contain 50 quasars in that bin. Its height shows the best-fit $\varepsilon$-value at that
				redshift. The average $\varepsilon$ value of all the subsamples is $0.629\pm 0.0873$.} 
			\label{gamma}
		\end{figure}
		
		\begin{figure*}
			\centering
			\includegraphics[width=0.75\textwidth]{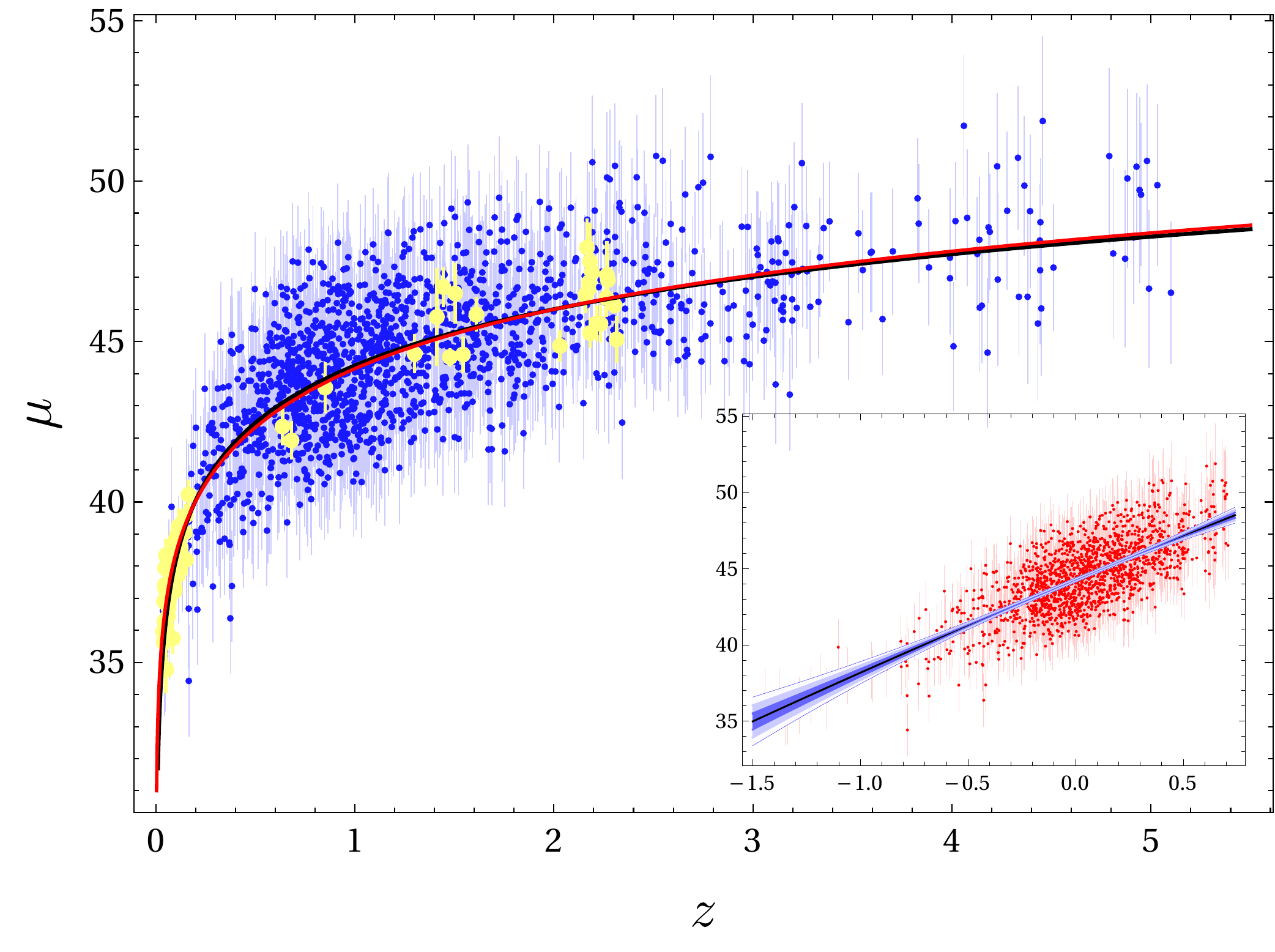}
			\caption{Distance modulus based on the \protect\cite{Risaliti_2019_Quasars} quasar
				sample. The blue data with errorbars are calibrated quasars, while the yellow points represent the 
				$\mathrm{H_{\uppercase\expandafter{\romannumeral2}}}$ galaxy measurements. The red curve shows the 
				{\it Planck}-optimized $\Lambda$CDM model, with $\Omega_m = 0.315\pm 0.007$ \citep{Planck_2018}. The 
				black curve is the best-fit $\mu(z)$ function, reconstructed with our B{\'e}zier method. The log 
				subplot provides more detail concerning the B{\'e}zier fit, including the optimized curve 
				and its 3-$\sigma$ uncertainty region.}
			\label{distance_modulus}
		\end{figure*}
		
		We follow \cite{Risaliti_2019_Quasars} in finding the slope $\varepsilon$ and the constant $\tilde{\alpha}$, though our approach is slightly modified. \cite{Risaliti_2019_Quasars} split the total quasar 
		sample into subsamples located in narrow redshift bins, within which $D_L$ in 
                Equation~\eqref{quasarlumi} may be considered to be constant if the bins are sufficiently small. In that case,
		the fluxes $F_\mathrm{X}$ and $F_\mathrm{UV}$ will have a similar log-linear relation to that between 
		$L_\mathrm{X}$ and $L_\mathrm{UV}$. \cite{Risaliti_2019_Quasars}  also found that the final value of 
		$\varepsilon$ is insensitive to the specified redshift bins as long as $\Delta\,\mathrm{log}(z)<0.1$.
		
			\begin{table}
			\renewcommand\arraystretch{1.2}
			{\caption{Best-fit parameters for all the models}}
			\label{table2}
			\begin{center}
				{\footnotesize
					\begin{tabular}{ccccc}
							\hline\hline
							Param. & SIS-$f_e$ 1   & SIS-$f_e$ 2 & SIS-$f_e$ 3 & 
							Non-SIS  \\
							\hline
							$\eta _0$ & $-0.021^{+0.068}_{-0.048}$ & $\cdots$ & $\cdots$ & 	
							$-0.109^{+0.044}_{-0.031}$\\
							$\eta _1$ & $\cdots$ & $-0.404^{+0.123}_{-0.090}$ & $\cdots$ & 
							$\cdots$ \\
							
							$\eta _2$ &$\cdots$ & $0.106^{+0.028}_{-0.034}$ & $\cdots$ & 
							$\cdots$ \\
							
							$\eta _3$ &$\cdots$ & $\cdots$ & $-0.507^{+0.193}_{-0.133}$ & 
							$\cdots$ \\
							
							$\nu$ & $0.043^{+0.039}_{-0.039}$& $0.016^{+0.043}_{-0.044}$ & 
							$0.031^{+0.043}_{-0.044}$ & 
							$0.065^{+0.040}_{-0.039}$ \\
							
							$f_e$ & $1.091^{+0.027}_{-0.026}$& $1.136^{+0.034}_{-0.032}$ & 
							$1.138^{+0.036}_{-0.034}$ & 
							$\cdots$ \\
							
							$\gamma$ & $\cdots$ & $\cdots$ & $\cdots$ & 
							$1.883^{+0.089}_{-0.091}$ \\
							
							$\delta$ & $\cdots$ & $\cdots$ & $\cdots$ & 
							$1.917^{+0.506}_{-0.433}$ \\
							
							$\beta$ &$\cdots$ & $\cdots$ & $\cdots$ & 
							$-0.572^{+0.793}_{-0.506}$ \\
							
							$\sigma_{\rm int}$ &$\le 0.048$ & $\le 0.046$& 
							$\le 0.047$ & $0.069^{+0.021}_{-0.024}$
							\\
							\hline	
						\end{tabular}
				}
			\end{center}
		\end{table} 
		
		In this paper, we split the quasar sample using the following procedure. First, we sort all 
		the quasars according to their redshift and make the first cut such that each subsample (except the last) 
		contains 50 quasars. This procedure makes full use of the data in quasar-dense redshift bins and 
		ensures a high-degree of accuracy. The best fit results for $\varepsilon$, from all the subsamples, 
		are shown in Figure~\ref{gamma}. We adopt the average value of all those individual subsample
		measurements, which equals $0.629\pm 0.0873$. In earlier work, one of us \citep{Fulvio_2019_quasar} used 
		the same data set, though with alternative approaches to finding $\varepsilon$. The three models 
		used to optimize these parameters in that work, including $\Lambda$CDM and the empirical cosmographic 
		model, straddle the value of $\varepsilon$ we have found here.

		\begin{figure}
			\centering
			\includegraphics[width=0.45\textwidth]{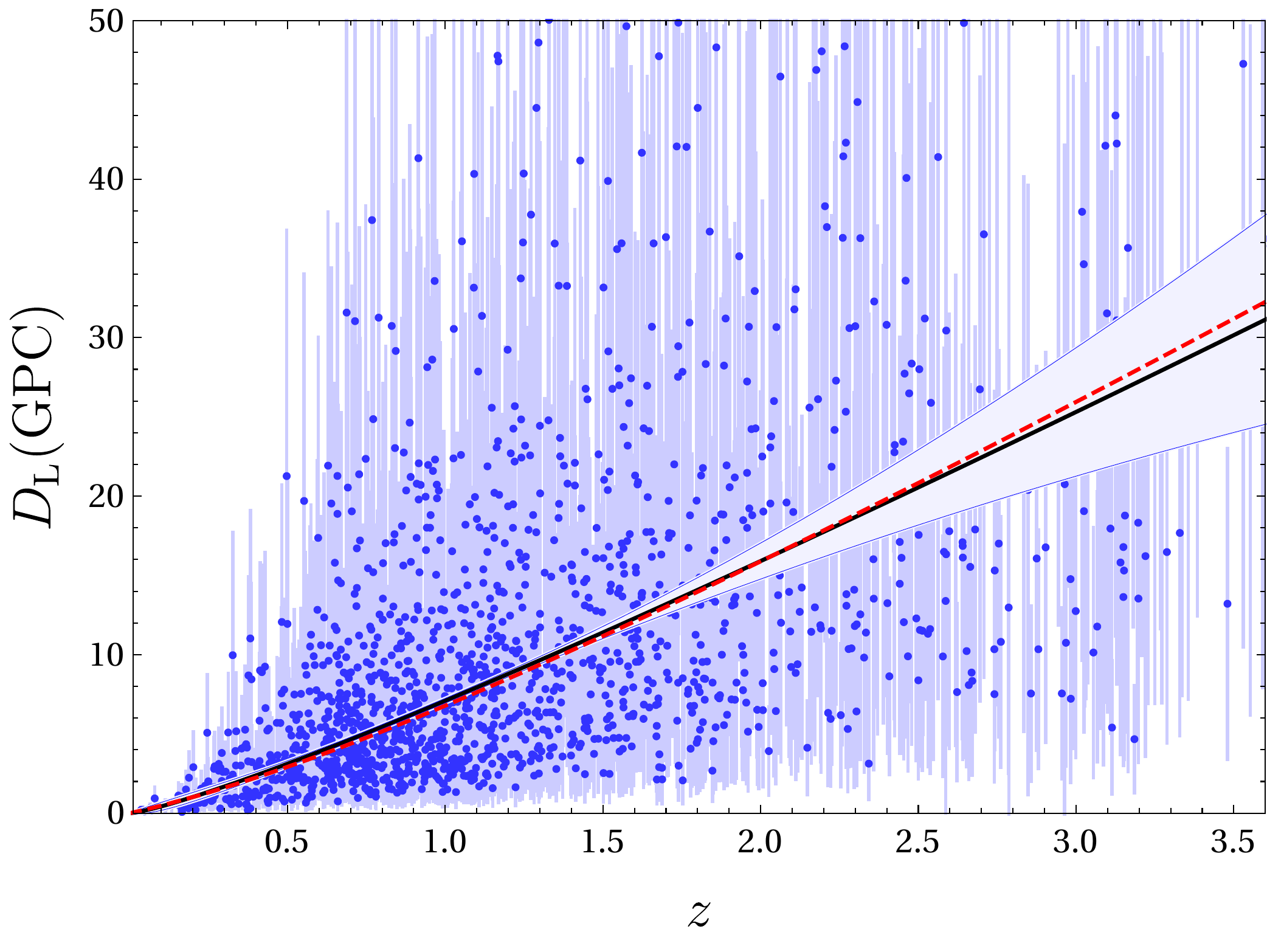}
			\caption{The luminosity distance calculated from the quasar data, and its 
				$1\sigma$ confidence region. The red dashed curve represents the theoretical luminosity distance 
				predicted by $\Lambda$CDM with $\Omega_m=0.315$ and a spatial curvature constant 
				$\Omega_K=0.001$ \citep{Planck_2018}. The theoretical curve lies well within the $1\sigma$ 
				confidence region.}
			\label{fig3}
		\end{figure}

		\begin{figure*}
			\centering
			\includegraphics[width=0.7\textwidth]{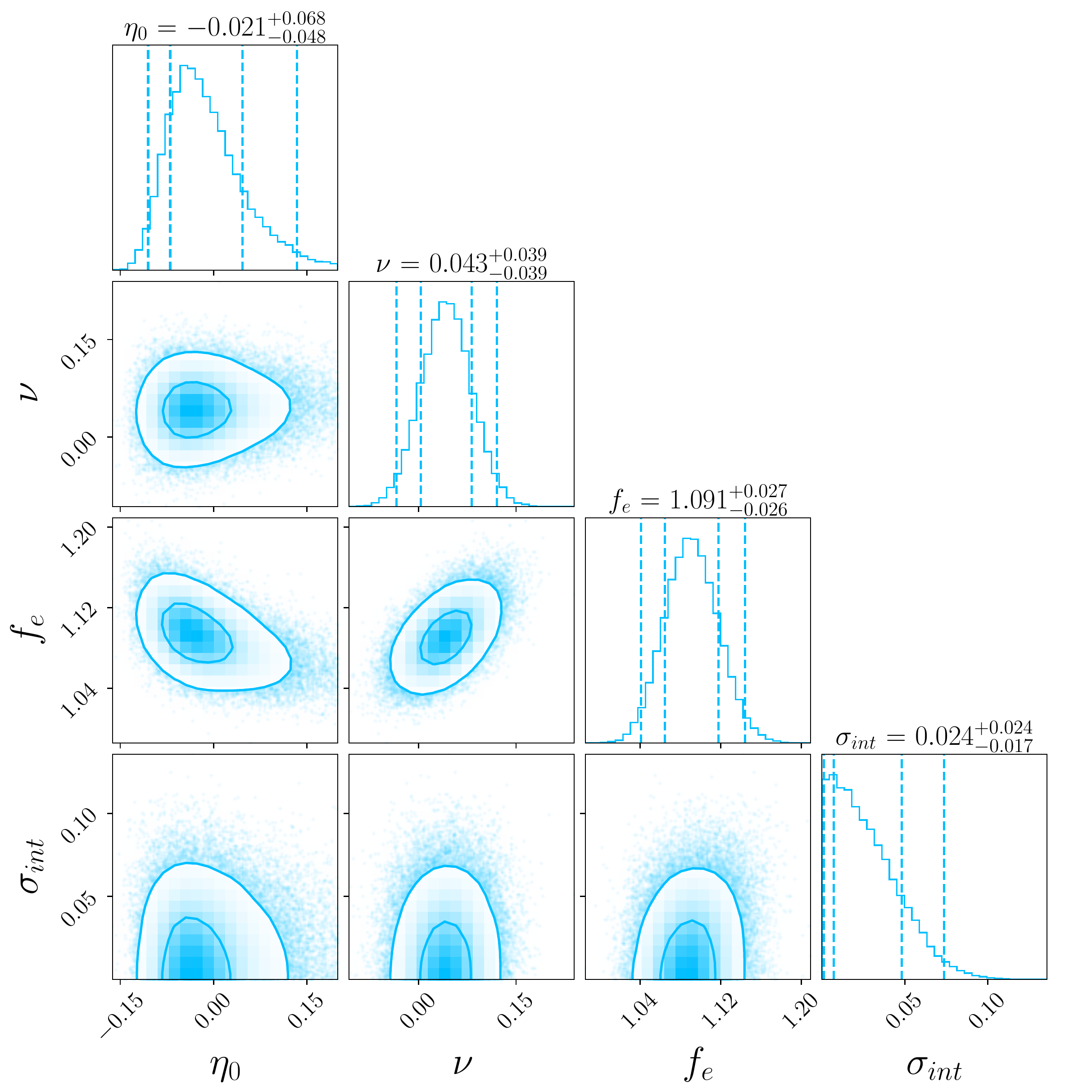}
			\caption{The 1D and 2D contours representing the $1 \sigma$ and $2 \sigma$ 
				confidence regions for the parameterization $\eta(z)=1+\eta_0z$ using the SIS 	lens model.}
			\label{eta_0_fe}
		\end{figure*}

		To find $\tilde{\alpha}$, we use the same method developed by \cite{Risaliti_2019_Quasars}, though calibrate the fits with $\mathrm{H_{\uppercase\expandafter{\romannumeral2}}}$ galaxies rather than 
		SNe. The luminosity ($L_\mathrm{H\beta}$) of Balmer lines from ionized hydrogen gas in these sources has a simple log-linear relationship with the velocity dispersion, $\sigma_\nu$, within the radiating plasma:
		\be
		\mathrm{log{\it L}_{H\beta}=\omega\, log\sigma_{\nu} +\tau}\;,
		\ee
		where $\omega$ and $\tau$ are two constants \citep{Melnick:1987,Melnick:1988,Fuentes-Masip:2000,Melnick:2000,Bosch:2002,Telles:2003,Siegel:2005,Bordalo:2011,Plionis:2011,Chavez:2012,Chavez:2014,Mania:2012,Terlevich:2015}. The distance modulus of an $\mathrm{H_{II}}$ galaxy can be written as
		\be
		\mu_{\mathrm{H\beta}}=-\tilde\tau+2.5(\omega\mathrm{log}\sigma_\nu- \mathrm{log}F_{H\beta})+25-5\mathrm{log}H_0\;,
		\label{muHII}
		\ee
		where $\tilde{\tau}=-2.5\tau-5\mathrm{log}H_0+125.2$ \citep{Wei_2016_HII}. Here $H_0$ is the Hubble constant and we set $H_{0}=67.35 \pm 16.47 \mathrm{~km} \mathrm{~s}^{-1} \mathrm{Mpc}^{-1}$, which is a model-independent value reconstructed by a machine learning method \citep{Wang_2020b}.
		In principle, the constants $\omega$ and $\tilde\tau$ should be constrained with a specified model.  These constants
		have been shown to vary indistinguishably between different expansion scenarios, however, 
so Equation \eqref{muHII} is effectively model-independent \citep{Wei_2016_HII,Ruan_2018_APJ_cdd}. 
This very weak dependence, if any, on the background cosmology is the principal reason 
we are opting to use $\mathrm{H_{\uppercase\expandafter{\romannumeral2}}}$ galaxies rather than SNe 
to calibrate the quasar sample. As described in the first paragraph of this section above, Type Ia
supernovae cannot be used to infer a true model-independent distance modulus, since the parameters
of the lightcurve must be optimized along with the model itself. Fortunately, the situation with
$\mathrm{H_{\uppercase\expandafter{\romannumeral2}}}$ galaxies is very different, because their use
thus far has shown that the constants in Equation~\eqref{muHII} are very insensitive to the
expansion rate. Following \cite{Ruan_2018_APJ_cdd}, we set $\omega= 4.87^{+0.11}_{-0.08}$ and 
$ \tilde\tau=32.42^{+0.42}_{-0.33}$ in this paper. By cross-matching the $\mathrm{H_{II}}$ data 
with quasars in the overlapping redshift range $(0.036, 2.315)$, we find an optimized value 
$\tilde{\alpha}=0.698$. 
		
		\begin{figure*}
		\centering
		\includegraphics[width=0.8\textwidth]{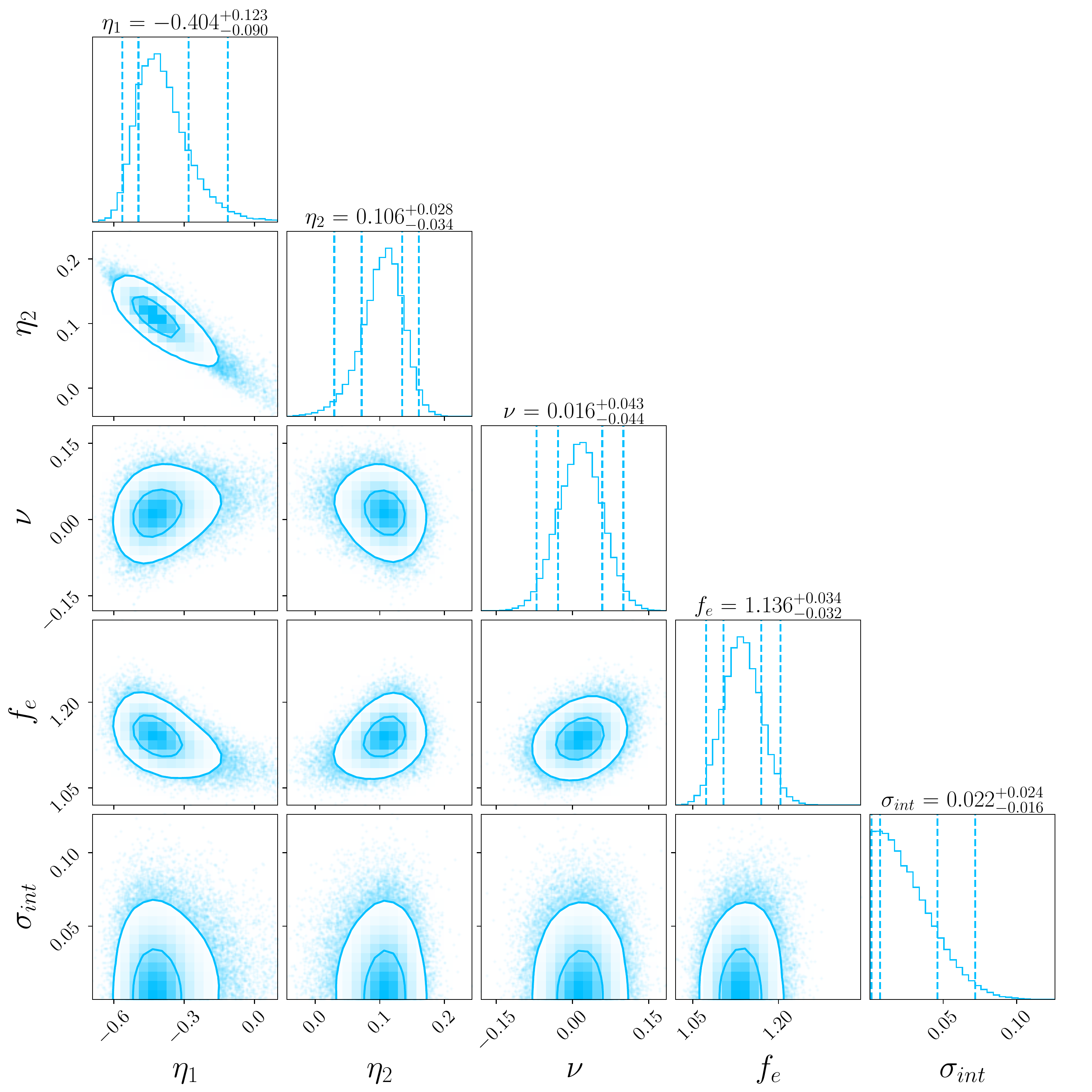}
		\caption{1D and 2D contours representing the $1 \sigma$ and $2 \sigma$ 
			confidence
			regions for the parameterization $\eta(z)=1+\eta_1z+\eta_2z^2$ using the
			SIS lens model.}
		\label{eta12_fe}
		\end{figure*}
	
		Finally, we use a B$\mathrm{\acute{e}}$zier 
		parametric fit to reconstruct the distance modulus, 
		\begin{equation}
			\mu_{n}(z)=\sum_{d=0}^{n} \xi_{d} h_{n}^{d}(z)\;, 
		\end{equation}
		where
		\begin{equation}
			h_{n}^{d}(z) \equiv \frac{n !\left(z /z_{m}\right)^{d}}
			{d !(n-d) !}\left(1-\frac{z}{z_{\mathrm{m}}}\right)^{n-d}\;,
		\end{equation}
		in terms of the maximum redshift $z_\mathrm{m}$ in the quasar data, and positive coefficients
		$\xi_{d}$. This method was first developed by \cite{Amati_2019_BezierMethod} to reconstruct 
		Hubble data and we use it here to reconstruct the continuous $\mu(z)$ function. Following 
		\cite{wei_2020_Chronometers}, we adopt n = 2 to fit the discretized distance-modulus data. 
		So the free parameters include: $\beta_0$, $\beta_1$, $\beta_2$, whose best-fit values are
		$44.255\pm0.046$, $59.258\pm0.354$ and $69.118\pm6.575$, respectively. The reconstructed 
		$\mu(z)$ function is shown in Figure~\ref{distance_modulus}.
		
		\begin{figure*}
		\centering
		\includegraphics[width=0.7\textwidth]{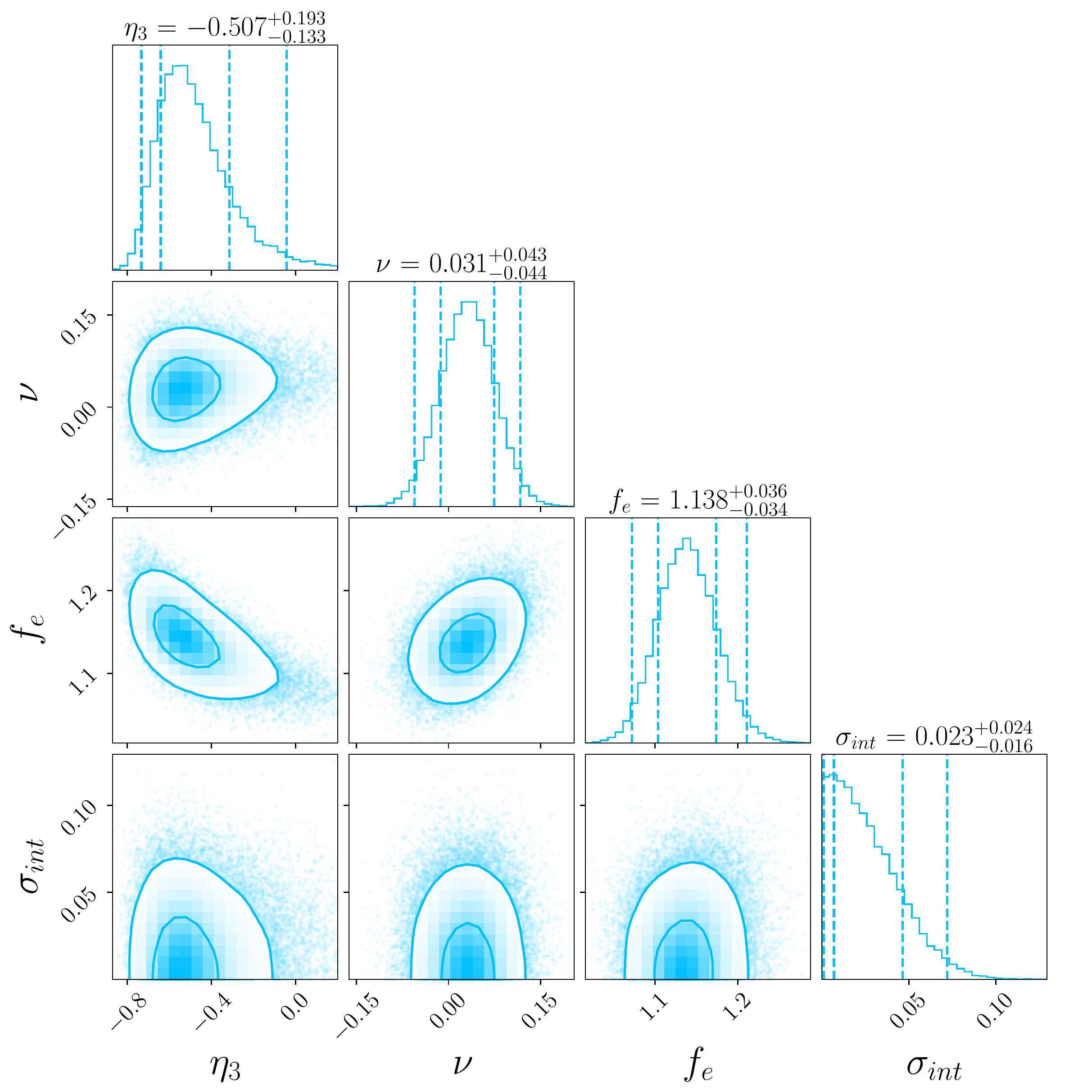}
		\caption{1D and 2D contours representing the $1 \sigma$ and $2 \sigma$ 
				confidence regions for the parameterization $\eta(z)=1+\eta_3z/(1+z)$ using the 
				SIS lens model.}
		\label{eta3_fe}
		\end{figure*}

		Given the relationship between $\mu(z)$ and $D_L$, we can directly obtain the dimensionless 
		luminosity distance of quasars as 
		\begin{equation}
			d_\mathrm{L}=10^{{\mu(z)}/{5}-5}/\mathrm{Mpc}\;,
		\end{equation}
		with a 1-$\sigma$ uncertainty
		\begin{equation}
			\sigma_{d_\mathrm{L}}=\frac{\mathrm{ln 10}}{5}d_\mathrm{L}\sigma_\mu\;.
		\end{equation}
	
		With the reconstructed luminosity distance, the uncertainty in $\mathscr{R}(d_L^{\,l},d_L^{\,s},z_l,z_s)$ 
		can be expressed as a function of $d_L^{\,l}$, $d_L^{\,s}$, $\sigma_{d_L}^{\,l}$ and $\sigma_{d_L}^{\,s}$:
	\be
	\sigma_{\mathscr{R}}=\sqrt{\left( K_l-\frac{1}{K_l}-
		\Phi K_s\right)^2\left(\frac{\sigma_{d_L^{\,l}}}{d_L^{\,l}}\right)^2+\left( 
		\frac{\Phi}{K_s}\right)^2\left(\frac{\sigma_{d_L^{\,s}}}{d_L^{\,s}}\right)^2}\;.
	\label{sigmaR}
	\ee
	Note that $K_l$ and $K_s$ equal one if $\Omega_K$ equals to zero, in which case 
	Equation~\eqref{sigmaR} simplifies to 
	\be
	\sigma_{\mathscr{R}}=\Phi\sqrt{\left(\frac{\sigma_{d_L^{\,l}}}{d_L^{\,l}}\right)^2+
		\left(\frac{\sigma_{d_L^s}}{d_L^s}\right)^2}\;.
	\label{sigma_R}
	\ee
	Equation~\eqref{sigma_R} has been used in previous work similar to that reported here.

	\section{RESULTS AND DISCUSSION}
	\subsection{The SIS model}
	In order to calculate the posterior distribution of the model parameters, we use 
	the Python module emcee\footnote[1]{\href[]{https://emcee.readthedocs.io/en/stable/}
		{https://emcee.readthedocs.io/en/stable/}}, which is an  Affine Invariant Markov chain 
	Monte Carlo (MCMC) Ensemble sampler \citep{emcee_2013}, to survey the posterior distribution 
	in parameter space and to maximize the likelihood function in Equation~\eqref{likehood}. 
	The resulting contour plots are made with the Python module 
	corner\footnote[2]{\href[]{https://corner.readthedocs.io/en/latest/}
		{https://corner.readthedocs.io/en/latest/}} \citep{corner}.
	
	We assumed the SIS lens model for the first application of the method described above.
	Combining the 109 SGL data with the $d_L$ function reconstructed 
		from the quasar measurements with the B{\'e}zier parametric method, we have found for
		the first parameterization $\eta(z)=1+\eta_0 z$ that
		\be
		\eta_0=-0.021^{+0.068}_{-0.048}\;,
		\ee 
		consistent with the CDDR to within 1 $\sigma$ accuracy. Our 1-$\sigma$ error is about half 
		the size of that found by \cite{Lyu_2020_apj_cdd}. Our result appears to favour a small 
		negative value, however. The posterior distribution of $\eta_0$, and that of other 
	related parameters, are plotted in Figure~\ref{eta_0_fe}. 
	
	For the more complex parameterization $\eta(z)=1+\eta_1z+\eta_2z^2$, the best-fit parameter 
	values are 
		\be
		\eta_1=-0.404^{+0.123}_{-0.090}\ \ {\rm and}\ \ \eta_2=0.106^{+0.028}_{-0.034}\;,
		\label{result_eta12}
		\ee
\noindent and the corresponding contour plots are shown in Figure~\ref{eta12_fe}. This second-order
	parameterization, which was also used previously by \cite{Ruan_2018_APJ_cdd}, allows for 
	greater precision. The 1-$\sigma$ errors in Equation~\eqref{result_eta12} are smaller than 
		those found by \cite{Ruan_2018_APJ_cdd}, a result stemming from the improved 
		capability of the B{\'e}zier method, as well as an increase in the SGL sample size.
	
	Finally, we have found for the third parameterization the optimized parameter value
		\be
		\eta_3=-\ 0.507^{\ +0.193}_{\ -0.133}\;,
		\ee
\noindent which disfavours zero by about $3\sigma$. This type of parameterization is more sensitive, so 
		it appears to deviate from zero more robustly than the first parameterization. The corresponding 
		contour plots are shown in Figure~\ref{eta3_fe}, and a summary of these results may be found in 
		Table~\ref{table2}.
	
	\subsection{The non-SIS model}
\begin{figure*}
		\centering
		\includegraphics[width=0.8\textwidth]{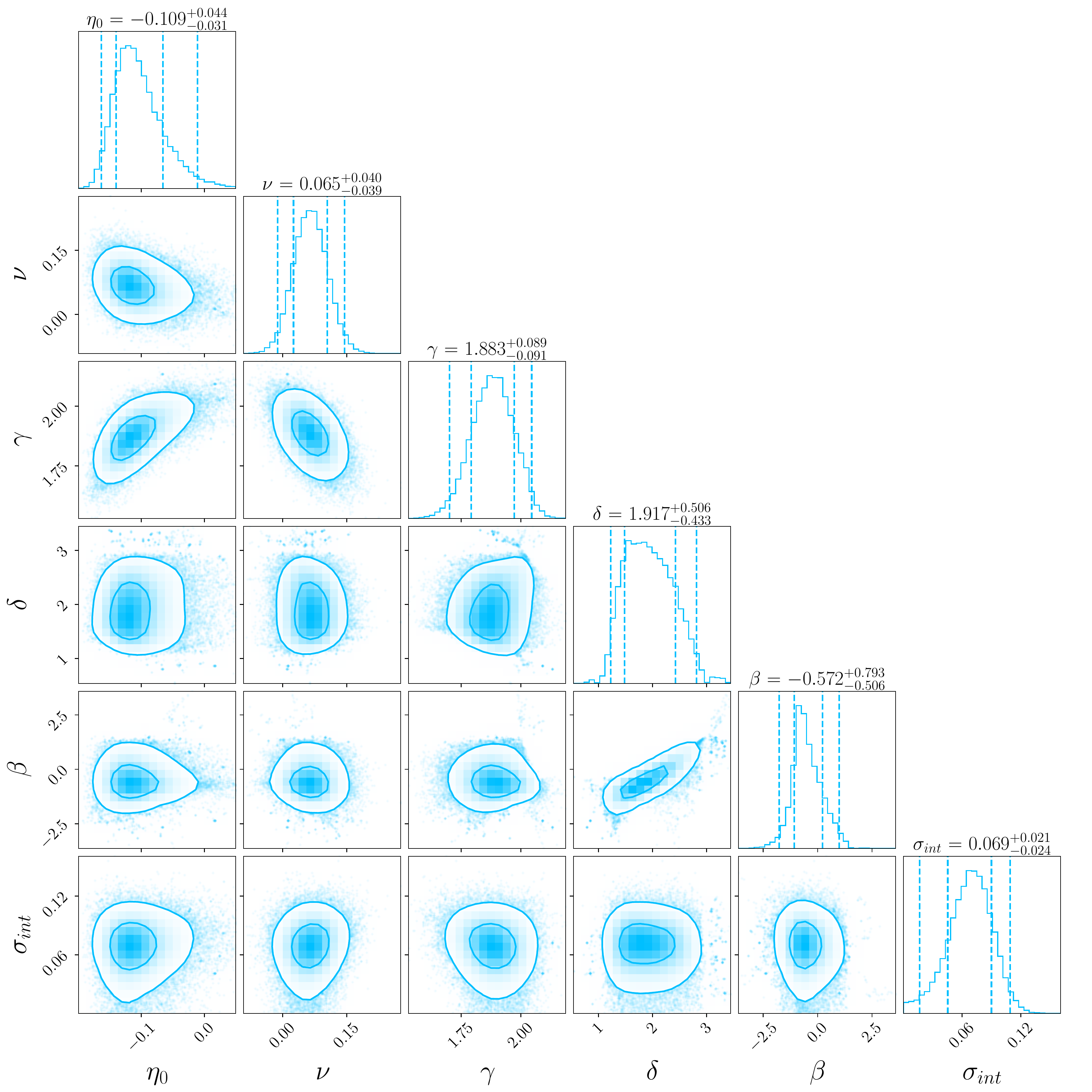}
		\caption{1D and 2D contours representing the $1\sigma$ and $2\sigma$ confidence 
		regions for the parameterization $\eta(z)=1+\eta_0z$ usnig a non-SIS lens model.}
		\label{eta0_non_SIS}
	\end{figure*}

	For the non-SIS lens model, we focus solely on the parameterization $\eta(z)=1+\eta_0 z$
	to avoid possible degeneracies among the larger number of free parameters (see Table~\ref{table2}).
	As it turns out, the result in this case is even stronger than that found for the SIS models,
	so the outcome is quite robust. The 1D and 2D marginalized distributions for $\eta_0$
	and the non-SIS parameters $\gamma$, $\delta$, $\beta$, etc. are shown in 
	Figure~\ref{eta0_non_SIS}. Specifically, the $1\sigma$ confidence limits are:\
		\be
		\begin{split}
			\eta_0 &= -0.109^{+0.044}_{-0.031}\\
			\gamma = 1.883^{+0.089}_{-0.091}\ \ \ \delta &= 1.917^{+0.506}_{-0.433}\ \ \ 
			\beta = -0.572^{+0.793}_{-0.506}
		\end{split}\;.
		\ee

	The non-SIS result for $\eta_0$ is even more precise than that of the SIS models, at least based 
	on the estimated errors. We have found that a zero value of $\eta_0$ is excluded
		at a confidence level $\sim 3\sigma$. As was the case for the SIS model, this constraint on 
		$\eta_0$ favours a small negative correction to the CDDR, based on the {\it Planck}-optimized
		$\Omega_K$ value. The remaining parameters in the non-SIS model are also optimized with high precision,
		with $\delta$ and $\beta$ consistent with their SIS values to within $1\sigma$.
	
	\subsection{The CDDR for more extreme deviations of $\Omega_K$ from zero}
	Although our focus thus far has been on the {\it Planck}-optimized value of 
		$\Omega_K$, our CDDR test may be applied to any other choice of $\Omega_K$. It is beyond
		the scope of the present paper, however, to thoroughly search parameter space in order
		to find an optimized value of $\Omega_K$ different from that found by {\it Planck}.
		Here, we demonstrate the outcome for the CDDR, based on a broad sample of $\Omega_K$ values. 
		
\begin{figure}
		\centering
		\includegraphics[width=0.45\textwidth]{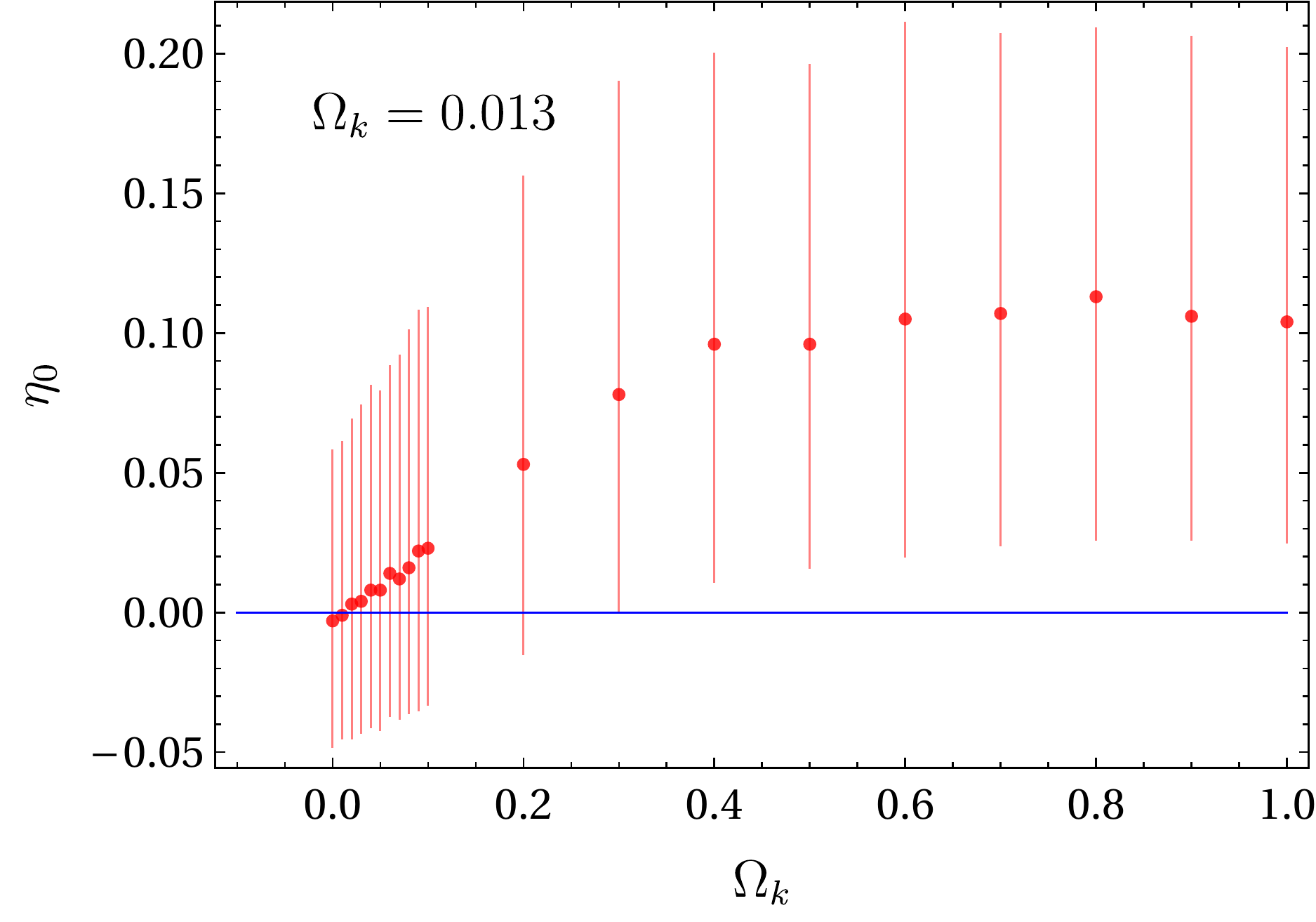}
		\caption{The relation between $\eta_0$ and $\Omega_K$ for a sample of
			spatial curvature values. A strict adherence to the CDDR, i.e., $\eta_0=0$, is associated
			with $\Omega_K=0.013$. The maximum deviation of $\eta_0$ from one appears to be $\sim 0.12$.}
		\label{omega_K}
	\end{figure}

		Generally speaking, the distance sum rule in Equation~\eqref{sumrule} is invalid for 
		$\Omega_K<0$ \citep{Hogg_1999_distance_measures}. We thus choose 20 positive $\Omega_K$ 
		values and repeat the above calculation using the SIS model with the parameterization 
		$1+\eta_0z$. The test results are shown in Figure~\ref{omega_K}, which clearly shows
		that the CDDR strongly favours an $\Omega_K$ near zero. Based on this limited survey, 
		however, the value of $\Omega_K$ consistent with the CDDR is actually somewhat
		positive, larger than that found by {\it Planck}. Though this is not yet strong evidence
		of a departure from the CDDR, the combination of results from {\it Planck} and our
		brief survey with other values of $\Omega_K$ is intriguing enough to warrant future
		consideration as the data continue to improve.	

\subsection{Possible factors that may bias the CDDR results}
We have made several efforts in this work to ensure that the results are reliable and robust. 
Nevertheless, there may still be other factors biasing our conclusions that could be fixed with
improved future observations, notably the SGL measurements. As noted, the SGL dataset used in 
this paper was assembled from six subsamples, each of which has its own set of systematic
errors. Therefore, our results may be biased due to possible inconsistencies in the SGL data. 
Unfortunately, we cannot rely on only one subset, since the greatly reduced number of sources
would imply even bigger errors. In previous work similar to ours, \cite{Lyu_2020_apj_cdd} also 
pointed out this issue and found that the CDDR tends to negative values. 

In addition to this,
there may be a possibility that an incorrect choice of $\Omega_K$ may be biasing the CDDR, but 
we cannot yet present a model-independent way of testing the value of $\Omega_K$ on its own. 
One needs to keep in mind that many previous model-independent tests of $\Omega_K$ were based
on the assumption that the CDDR is valid. Of course, any model-independent and CDDR-independent 
test of $\Omega_K$ would provide more robust conclusions regarding the CDDR itself. Looking to
the future, our high-precision method of testing the CDDR should make deviations from Equation~(1) 
more obvious if the spatial curvature in the real Universe turns out to be different from zero.
	
\vfill
	\section{Conclusion}
	In this paper, we have demonstrated that the CDDR may be tested in a 
		model-independent way by combining the luminosity distance inferred from high-redshift 
		quasars with the angular diameter distance obtained from SGL's. This approach avoids 
	potential difficulties faced by previously used methodologies based on Type Ia SNe. It 
	is widely known that one needs to assume a specific cosmological model when using SNe 
	in order to optimize the so-called ``nuisance" parameters, rendering all such approaches 
	model-dependent. Under such circumstances, one could not rule out the possibility that a
	deviation of $\eta_0$ from $0$ is caused by the incorrect model, rather than the data 
	themselves.
	
	Fortunately, we now have a much broader array of cosmological measurements
	offering alternatives to the use of Type Ia SNe, avoiding possible weaknesses
	stemming from the need to pre-assume some particular model. In this paper,
	we have proposed one such method and, in addition, have advanced the analysis
	by another significant step, i.e., by avoiding the need to assume a spatially
	flat background. Instead, we have carried out our analysis for 
		several different $\Omega_K$ values, including the optimization 
		$\Omega_K=0.001$ from {\it Planck} \citep{Planck_2018}.
	
	Our analysis has benefitted from the use of the B{\'e}zier reconstruction 
		technique which, as we have seen, produces a more precise optimization of model parameters.
		With the smaller errors found in our results, our analysis has shown that {\it Planck}-$\Lambda$CDM 
		(with $\Omega_K=0.001$) is somewhat valid with the CDDR, though a slightly negative correction
		is favoured by these data. This approach avoids the introduction of possible systematic errors 
		associated with the cosmological models, so any possible violation of the CDDR would originate 
		from the physics itself.
		
		By carrying out our analysis without assuming zero spatial curvature, this approach
		has extended the range of previous CDDR tests. There exists an abundance of evidence 
		suggesting that $\Omega_K$ is probably zero, including the results from {\it Planck}, but
		we have allowed the widest possible range of outcomes by allowing a deviation from
		complete spatial flatness. We have therefore probed the viability of the CDDR more
		generally, and have found that---at worst---any deviation of $\Omega_K$ from zero
		consistent with the CDDR is much smaller than one. Nevertheless, the violation of the
		CDDR---should $\Omega_K$ be consistent with the value found by {\it Planck}, or even
		slightly larger---represents some tension at the level of $\sim 2$-$\sigma$ when using 
		the parameterization $\eta(z)=1+\eta_0z$.
	
	Looking to the future, an increase in the number of SGL measurements
		would improve the precision of our approach even further. But already with the 161 data 
		points at hand, combined with the B{\'e}zier method of reconstructing the luminosity distance, 
		we have reduced the errors found in \cite{Ruan_2018_APJ_cdd} by about a factor $1.5$ for
		the SIS model, and a factor $3-4$ for the non-SIS lens model. As the accuracy of such 
		measurements continues to improve, we anticipate ruling out a non-flat cosmology with 
		even greater confidence. Of course, the alternative could also be a clearer indication 
		of a violation of the CDDR, requiring the introduction of new physics.
	
	\section*{acknowledgments}
        We are very grateful to the anonymous referee for comments that have led
        to a significant improvement of the analysis and presentation in this paper.
	FM is grateful to Amherst College for its support through a John Woodruff 
	Simpson Lectureship. This work was partially supported by the National Key R \& D Program of China (2017YFA0402600) and the National Science Foundation of China (Grants No.11929301, 11573006).

	
	\section*{DATA AVAILABILITY STATEMENT}
		All of the SGL data can be taken from \cite{Chen_2019MNRAS_NewLensData} and the
		high-redshift quasar sample is available from the corresponding author of the 
		paper by \cite{Risaliti_2019_Quasars}. The $\mathrm{H_{II}}$ data used to calibrate 
                the quasar distance modulus may be obtained from \cite{Wei_2016_HII}.
	
	\bibliographystyle{apj}
	\bibliography{ms}

\begin{thebibliography}{}
\expandafter\ifx\csname natexlab\endcsname\relax\def\natexlab#1{#1}\fi

\bibitem[{{Amati} {et~al.}(2019){Amati}, {D'Agostino}, {Luongo}, {Muccino}, \&
  {Tantalo}}]{Amati_2019_BezierMethod}
{Amati}, L., {D'Agostino}, R., {Luongo}, O., {Muccino}, M., \& {Tantalo}, M.
  2019, \mnras, 486, L46

\bibitem[{{Bassett} \& {Kunz}(2004)}]{Bassett_2004_cddbeigins}
{Bassett}, B.~A., \& {Kunz}, M. 2004,
  \href{https://doi.org/10.1103/PhysRevD.69.101305}{\prd}, 69, 101305

\bibitem[{{Bernstein}(2006)}]{Bernstein_2006}
{Bernstein}, G. 2006, \apj, 637, 598

\bibitem[{{Bordalo} \& {Telles}(2011)}]{Bordalo:2011}
{Bordalo}, V., \& {Telles}, E. 2011, \apj, 735, 52

\bibitem[{{Bosch} {et~al.}(2002){Bosch}, {Terlevich}, \&
  {Terlevich}}]{Bosch:2002}
{Bosch}, G., {Terlevich}, E., \& {Terlevich}, R. 2002, \mnras, 329, 481

\bibitem[{Cao {et~al.}(2016)Cao, Biesiada, Yao, \& Zhu}]{cao_2016_mnras}
Cao, S., Biesiada, M., Yao, M., \& Zhu, Z.-H. 2016, Monthly Notices of the
  Royal Astronomical Society, 461, 2192

\bibitem[{{Ch{\'a}vez} {et~al.}(2012){Ch{\'a}vez}, {Terlevich}, {Terlevich},
  {Plionis}, {Bresolin}, {Basilakos}, \& {Melnick}}]{Chavez:2012}
{Ch{\'a}vez}, R., {Terlevich}, E., {Terlevich}, R., {et~al.} 2012, \mnras, 425,
  L56

\bibitem[{{Ch{\'a}vez} {et~al.}(2014){Ch{\'a}vez}, {Terlevich}, {Terlevich},
  {Bresolin}, {Melnick}, {Plionis}, \& {Basilakos}}]{Chavez:2014}
{Ch{\'a}vez}, R., {Terlevich}, R., {Terlevich}, E., {et~al.} 2014, \mnras, 442,
  3565

\bibitem[{{Chen} {et~al.}(2019){Chen}, {Li}, {Shu}, \&
  {Cao}}]{Chen_2019MNRAS_NewLensData}
{Chen}, Y., {Li}, R., {Shu}, Y., \& {Cao}, X. 2019,
  \href{https://doi.org/10.1093/mnras/stz1902}{\mnras}, 488, 3745

\bibitem[{De~Bernardis {et~al.}(2006)De~Bernardis, Giusarma, \&
  Melchiorri}]{Bernardis_2006}
De~Bernardis, F., Giusarma, E., \& Melchiorri, A. 2006,
  \href{https://doi.org/10.1142/S0218271806008486}{IJMPD}, 15, 759

\bibitem[{Ellis {et~al.}(2013)Ellis, Poltis, Uzan, \& Weltman}]{Ellis_2013}
Ellis, G. F.~R., Poltis, R., Uzan, J.-P., \& Weltman, A. 2013,
  \href{https://doi.org/10.1103/PhysRevD.87.103530}{\prl}, 87, 103530

\bibitem[{{Etherington}(1933)}]{Etherington_1993}
{Etherington}, I.~M.~H. 1933, Philosophical Magazine, 15, 761

\bibitem[{{Etherington}(2007)}]{Etherington_2007}
---. 2007, General Relativity and Gravitation, 39, 1055

\bibitem[{Foreman-Mackey(2016)}]{corner}
Foreman-Mackey, D. 2016, \href{https://doi.org/10.21105/joss.00024}{The Journal
  of Open Source Software}, 1, 24

\bibitem[{Foreman-Mackey {et~al.}(2013)Foreman-Mackey, Hogg, Lang, \&
  Goodman}]{emcee_2013}
Foreman-Mackey, D., Hogg, D.~W., Lang, D., \& Goodman, J. 2013, Publications of
  the Astronomical Society of the Pacific, 125, 306

\bibitem[{{Fuentes-Masip} {et~al.}(2000){Fuentes-Masip},
  {Mu{\~n}oz-Tu{\~n}{\'o}n}, {Casta{\~n}eda}, \&
  {Tenorio-Tagle}}]{Fuentes-Masip:2000}
{Fuentes-Masip}, O., {Mu{\~n}oz-Tu{\~n}{\'o}n}, C., {Casta{\~n}eda}, H.~O., \&
  {Tenorio-Tagle}, G. 2000, \apj, 120, 752

\bibitem[{{Hogg}(1999)}]{Hogg_1999_distance_measures}
{Hogg}, D.~W. 1999, arXiv e-prints, astro

\bibitem[{Holanda {et~al.}(2010)Holanda, Lima, \& Ribeiro}]{Holanda_2010_APJL}
Holanda, R. F.~L., Lima, J. A.~S., \& Ribeiro, M.~B. 2010,
  \href{https://doi.org/10.1088/2041-8205/722/2/l233}{\apjl}, 722, L233

\bibitem[{Jones {et~al.}(2013)Jones, Rodney, Riess, Mobasher, Dahlen, McCully,
  Frederiksen, Casertano, Hjorth, Keeton, Koekemoer, Strolger, Wiklind,
  Challis, Graur, Hayden, Patel, Weiner, Filippenko, Garnavich, Jha, Kirshner,
  Ferguson, Grogin, \& Kocevski}]{Jones_2013}
Jones, D.~O., Rodney, S.~A., Riess, A.~G., {et~al.} 2013, The Astrophysical
  Journal, 768, 166

\bibitem[{Khedekar \& Chakraborti(2011)}]{Khedekar_2011_PRL_cdd21cm}
Khedekar, S., \& Chakraborti, S. 2011,
  \href{https://doi.org/10.1103/PhysRevLett.106.221301}{\prl}, 106, 221301

\bibitem[{{Li} \& {Lin}(2018)}]{Li_2018_MNRAS_cdd}
{Li}, X., \& {Lin}, H.-N. 2018, \mnras, 474, 313

\bibitem[{Li {et~al.}(2011)Li, Wu, \& Yu}]{Li_2011_APJL_cdd}
Li, Z., Wu, P., \& Yu, H. 2011,
  \href{https://doi.org/10.1088/2041-8205/729/1/L14}{\apjl}, 729, L14

\bibitem[{{Liao} {et~al.}(2016){Liao}, {Li}, {Cao}, {Biesiada}, {Zheng}, \&
  {Zhu}}]{Liao_2016ApJ_cdd}
{Liao}, K., {Li}, Z., {Cao}, S., {et~al.} 2016, \apj, 822, 74

\bibitem[{{Lin} {et~al.}(2018){Lin}, {Li}, \& {Li}}]{Lin_2018_MNRAS_cdd}
{Lin}, H.-N., {Li}, M.-H., \& {Li}, X. 2018, \mnras, 480, 3117

\bibitem[{{Lv} \& {Xia}(2016)}]{Lv_2016PDU_cdd}
{Lv}, M.-Z., \& {Xia}, J.-Q. 2016, Physics of the Dark Universe, 13, 139

\bibitem[{{Lyu} {et~al.}(2020){Lyu}, {Li}, \& {Xia}}]{Lyu_2020_apj_cdd}
{Lyu}, M.-Z., {Li}, Z.-X., \& {Xia}, J.-Q. 2020, \apj, 888, 32

\bibitem[{{Mania} \& {Ratra}(2012)}]{Mania:2012}
{Mania}, D., \& {Ratra}, B. 2012, Physics Letters B, 715, 9

\bibitem[{{Melia}(2019)}]{Fulvio_2019_quasar}
{Melia}, F. 2019, \mnras, 489, 517

\bibitem[{{Melnick} {et~al.}(1987){Melnick}, {Moles}, {Terlevich}, \&
  {Garcia-Pelayo}}]{Melnick:1987}
{Melnick}, J., {Moles}, M., {Terlevich}, R., \& {Garcia-Pelayo}, J.-M. 1987,
  \mnras, 226, 849

\bibitem[{{Melnick} {et~al.}(1988){Melnick}, {Terlevich}, \&
  {Moles}}]{Melnick:1988}
{Melnick}, J., {Terlevich}, R., \& {Moles}, M. 1988, \mnras, 235, 297

\bibitem[{{Melnick} {et~al.}(2000){Melnick}, {Terlevich}, \&
  {Terlevich}}]{Melnick:2000}
{Melnick}, J., {Terlevich}, R., \& {Terlevich}, E. 2000, \mnras, 311, 629

\bibitem[{Meng {et~al.}(2012)Meng, Zhang, Zhan, \& Wang}]{Meng_2012_cdd}
Meng, X.-L., Zhang, T.-J., Zhan, H., \& Wang, X. 2012,
  \href{https://doi.org/10.1088/0004-637X/745/1/98}{\apj}, 745, 98

\bibitem[{Nair {et~al.}(2011)Nair, Jhingan, \& Jain}]{Nair_2011_cdd}
Nair, R., Jhingan, S., \& Jain, D. 2011,
  \href{https://doi.org/10.1088/1475-7516/2011/05/023}{\jcap}, 2011, 023

\bibitem[{Peebles(1993)}]{peebles1993principles}
Peebles, P. J.~E. 1993, Principles of physical cosmology (Princeton university
  press)

\bibitem[{{Planck Collaboration} {et~al.}(2018){Planck Collaboration},
  {Aghanim}, {Akrami}, {Ashdown}, {Aumont}, {Baccigalupi}, {Ballardini},
  {Banday}, {Barreiro}, {Bartolo}, {Basak}, {Battye}, {Benabed}, {Bernard},
  {Bersanelli}, {Bielewicz}, {Bock}, {Bond}, {Borrill}, {Bouchet}, {Boulanger},
  {Bucher}, {Burigana}, {Butler}, {Calabrese}, {Cardoso}, {Carron},
  {Challinor}, {Chiang}, {Chluba}, {Colombo}, {Combet}, {Contreras}, {Crill},
  {Cuttaia}, {de Bernardis}, {de Zotti}, {Delabrouille}, {Delouis}, {Di
  Valentino}, {Diego}, {Dor{\'e}}, {Douspis}, {Ducout}, {Dupac}, {Dusini},
  {Efstathiou}, {Elsner}, {En{\ss}lin}, {Eriksen}, {Fantaye}, {Farhang},
  {Fergusson}, {Fernandez-Cobos}, {Finelli}, {Forastieri}, {Frailis},
  {Fraisse}, {Franceschi}, {Frolov}, {Galeotta}, {Galli}, {Ganga},
  {G{\'e}nova-Santos}, {Gerbino}, {Ghosh}, {Gonz{\'a}lez-Nuevo}, {G{\'o}rski},
  {Gratton}, {Gruppuso}, {Gudmundsson}, {Hamann}, {Handley}, {Hansen},
  {Herranz}, {Hildebrandt}, {Hivon}, {Huang}, {Jaffe}, {Jones}, {Karakci},
  {Keih{\"a}nen}, {Keskitalo}, {Kiiveri}, {Kim}, {Kisner}, {Knox},
  {Krachmalnicoff}, {Kunz}, {Kurki-Suonio}, {Lagache}, {Lamarre}, {Lasenby},
  {Lattanzi}, {Lawrence}, {Le Jeune}, {Lemos}, {Lesgourgues}, {Levrier},
  {Lewis}, {Liguori}, {Lilje}, {Lilley}, {Lindholm}, {L{\'o}pez-Caniego},
  {Lubin}, {Ma}, {Mac{\'\i}as-P{\'e}rez}, {Maggio}, {Maino}, {Mandolesi},
  {Mangilli}, {Marcos-Caballero}, {Maris}, {Martin}, {Martinelli},
  {Mart{\'\i}nez-Gonz{\'a}lez}, {Matarrese}, {Mauri}, {McEwen}, {Meinhold},
  {Melchiorri}, {Mennella}, {Migliaccio}, {Millea}, {Mitra},
  {Miville-Desch{\^e}nes}, {Molinari}, {Montier}, {Morgante}, {Moss}, {Natoli},
  {N{\o}rgaard-Nielsen}, {Pagano}, {Paoletti}, {Partridge}, {Patanchon},
  {Peiris}, {Perrotta}, {Pettorino}, {Piacentini}, {Polastri}, {Polenta},
  {Puget}, {Rachen}, {Reinecke}, {Remazeilles}, {Renzi}, {Rocha}, {Rosset},
  {Roudier}, {Rubi{\~n}o-Mart{\'\i}n}, {Ruiz-Granados}, {Salvati}, {Sandri},
  {Savelainen}, {Scott}, {Shellard}, {Sirignano}, {Sirri}, {Spencer},
  {Sunyaev}, {Suur-Uski}, {Tauber}, {Tavagnacco}, {Tenti}, {Toffolatti},
  {Tomasi}, {Trombetti}, {Valenziano}, {Valiviita}, {Van Tent}, {Vibert},
  {Vielva}, {Villa}, {Vittorio}, {Wand elt}, {Wehus}, {White}, {White},
  {Zacchei}, \& {Zonca}}]{Planck_2018}
{Planck Collaboration}, {Aghanim}, N., {Akrami}, Y., {et~al.} 2018, arXiv
  e-prints, \href{https://arxiv.org/abs/1807.06209}{arXiv:1807.06209}

\bibitem[{{Plionis} {et~al.}(2011){Plionis}, {Terlevich}, {Basilakos},
  {Bresolin}, {Terlevich}, {Melnick}, \& {Chavez}}]{Plionis:2011}
{Plionis}, M., {Terlevich}, R., {Basilakos}, S., {et~al.} 2011, \mnras, 416,
  2981

\bibitem[{{R{\"a}s{\"a}nen} {et~al.}(2015){R{\"a}s{\"a}nen}, {Bolejko}, \&
  {Finoguenov}}]{Rasanen_2015PRL}
{R{\"a}s{\"a}nen}, S., {Bolejko}, K., \& {Finoguenov}, A. 2015, \prl, 115,
  101301

\bibitem[{{Risaliti} \& {Lusso}(2019)}]{Risaliti_2019_Quasars}
{Risaliti}, G., \& {Lusso}, E. 2019, Nature Astronomy, 3, 272

\bibitem[{{Ruan} {et~al.}(2018){Ruan}, {Melia}, \& {Zhang}}]{Ruan_2018_APJ_cdd}
{Ruan}, C.-Z., {Melia}, F., \& {Zhang}, T.-J. 2018, \apj, 866, 31

\bibitem[{{Siegel} {et~al.}(2005){Siegel}, {Guzm{\'a}n}, {Gallego}, {Ordu{\~n}a
  L{\'o}pez}, \& {Rodr{\'\i}guez Hidalgo}}]{Siegel:2005}
{Siegel}, E.~R., {Guzm{\'a}n}, R., {Gallego}, J.~P., {Ordu{\~n}a L{\'o}pez},
  M., \& {Rodr{\'\i}guez Hidalgo}, P. 2005, \mnras, 356, 1117

\bibitem[{{Telles}(2003)}]{Telles:2003}
{Telles}, E. 2003, in Astronomical Society of the Pacific Conference Series,
  Vol. 297, Star Formation Through Time, ed. E.~{Perez}, R.~M. {Gonzalez
  Delgado}, \& G.~{Tenorio-Tagle}, 143

\bibitem[{{Terlevich} {et~al.}(2015){Terlevich}, {Terlevich}, {Melnick},
  {Ch{\'a}vez}, {Plionis}, {Bresolin}, \& {Basilakos}}]{Terlevich:2015}
{Terlevich}, R., {Terlevich}, E., {Melnick}, J., {et~al.} 2015, \mnras, 451,
  3001

\bibitem[{{Uzan} {et~al.}(2004){Uzan}, {Aghanim}, \&
  {Mellier}}]{Uzan_2004_prd_cdd}
{Uzan}, J.-P., {Aghanim}, N., \& {Mellier}, Y. 2004,
  \href{https://doi.org/10.1103/PhysRevD.70.083533}{\prd}, 70, 083533

\bibitem[{{Wang} {et~al.}(2020){Wang}, {Ma}, \& {Xia}}]{Wang_2020b}
{Wang}, G.-J., {Ma}, X.-J., \& {Xia}, J.-Q. 2020, arXiv e-prints,
  arXiv:2004.13913

\bibitem[{{Wei} \& {Melia}(2020)}]{wei_2020_Chronometers}
{Wei}, J.-J., \& {Melia}, F. 2020, \apj, 888, 99

\bibitem[{{Wei} {et~al.}(2016){Wei}, {Wu}, \& {Melia}}]{Wei_2016_HII}
{Wei}, J.-J., {Wu}, X.-F., \& {Melia}, F. 2016, \mnras, 463, 1144

\end{thebibliography}
	
	\label{lastpage}
	
\end{document}